\documentclass[]{aa} 
\usepackage{graphicx,natbib}
\usepackage{txfonts}
\begin{document}
   \title{Millimetre continuum observations of comet C/2009 P1 (Garradd)\thanks{Based on observations carried out with the IRAM Plateau de Bure Interferometer. IRAM is supported by INSU/CNRS (France), MPG (Germany), and IGN (Spain).}}

   \author{J.~Boissier \inst{1,2}
          \and
          D.~Bockel\'ee-Morvan \inst{3}
          \and
          O.~Groussin \inst{4}
          \and
          P.~Lamy \inst{4}
          \and
          N.~Biver \inst{3}
          \and
          J.~Crovisier \inst{3}
          \and
          P.~Colom \inst{3}
          \and
          R.~Moreno \inst{3}
          \and
          L.~Jorda \inst{4}
          \and
          V.~Pi\'etu \inst{5}}

   \institute{Istituto di Radioastronomia - INAF, Via Gobetti 101, Bologna, Italy (e-mail: boissier@ira.inaf.it)
   \and
   ESO, Karl Schwarzschild Str. 2, 85748 Garching bei Muenchen, Germany
   \and
   LESIA -- Observatoire de Paris, CNRS, UPMC, Universit\'e Paris-Diderot, 5 place Jules Janssen, 92195 Meudon, France.
   \and
   Aix Marseille Universit\'e, CNRS, LAM (Laboratoire d'Astrophysique de 
   Marseille) UMR 7326, 13388, Marseille, France
   \and
   Institut de radioastronomie millim\'etrique, 300 Rue de la Piscine, 38406 Saint Martin d'H\`eres, France
}

   \date{Received September XXX; accepted XXX}

 
  \abstract
   {Little is known about the physical properties of the nuclei of Oort cloud comets. Measuring the thermal emission of a nucleus is one of
the few means for deriving its size
and constraining some of its thermal properties. 
}
   {We attempted  to measure the nucleus size  of the Oort cloud comet C/2009~P1 (Garradd).
}
   { We used the Plateau de Bure Interferometer to measure the millimetric thermal emission of this comet  at 157~GHz (1.9~mm) and 266~GHz (1.1~mm).
}
   {Whereas the observations at 266~GHz were not usable due to bad atmospheric conditions, 
we derived a 3$\sigma$ upper limit on the comet continuum emission of 0.41~mJy at 157~GHz.
Using a  thermal model for a spherical nucleus with  standard thermal parameters, we found an upper limit of 5.6~km for the radius.
The dust contribution to our signal is estimated to be negligible.
Given the water production rates measured for this comet and our upper limit, we estimated that Garradd was very active, with an active fraction of its nucleus  larger than 50\%.
}
 {}


   \maketitle
%

\section{Introduction}
Cometary nuclei are among the most primitive objects in the Solar System, and their physical and chemical properties are thought to preserve a record of the  conditions that existed in the early solar nebula. 
Dynamical arguments support the hypothesis that they originate from at least two different regions. 
Following the dynamics-based classification of \cite{lev96}, ecliptic comets (ECs) are thought to originate in the Kuiper belt \citep{fer80}, thus explaining their low inclinations and prograde orbits. 
In contrast, nearly isotropic comets (NICs) are thought to have formed in the giant planets' region. 
They were members of a population of planetesimals that were scattered by planetary perturbations to the outskirts of the Solar System 3.5--4.5~Gyr ago, where their orbits were subsequently isotropized by gravitational perturbations from nearby stars and molecular clouds to form the roughly spheroidal Oort cloud. 
Some of the NICs were later perturbed again by passing stars, molecular clouds, or galactic tide, and their perihelia lowered to within $\sim$3~AU, where they became active and could be detected. 
The detected NICs are either ``returning'' NICs on elliptical orbits, or ``new'' NICs, presumably on their first passage through the inner Solar System.

\cite{lam+04} completed a review of the properties of cometary nuclei and concluded that there are reliable size determinations for only 13 NICs. The range of radii is surprisingly broad, 0.4 to 37~km, much broader than that of the ECs (0.6 to 5~km for $>$98\% of them). 
The NIC cumulative  size  distribution (CSD) substantially  differs from that of a  collisionally relaxed  population \citep{doh69,obrigreen03}, but with only 13 objects, a robust conclusion cannot be drawn.

 Measuring the millimetric thermal emission of a nucleus is 
way to estimate its size which depends moderately on the assumption made on its albedo.
However, due to the faintness of the signal, such observations are challenging.
Up to now, this could only be achieved for  C/1995 O1 (Hale-Bopp) and 8P/Tuttle \citep{alt+99,boi+11}. 

Comet C/2009 P1 (Garradd), referred to as Garradd hereafter, was discovered in 2009, when it was at 8.7~AU from the Sun \citep{mcngar09}.
With an eccentricity of 1 and an inclination of 106$^\circ$, Garradd undoubtedly belongs to the NIC class.
It  presented a high  level of activity  during its approach to the Sun  \citep[e.g., a water production rate of $1-2 \times 10^{29}$ s$^{-1}$ at 1.9~AU  was measured with the Herschel telescope in 2011 by][]{boc+12}.
IR and millimetric observations revealed that Garradd is CO-rich with a CO/H$_2$O abundance of the order of 10\% \citep{pag+12,biv+12dps}.

We attempted  to measure the nucleus size  of comet C/2009~P1 (Garradd).
We present here the observations of its thermal emission  carried out at millimetre wavelengths with the IRAM Plateau de Bure Interferometer (hereafter PdBI) in March 2012.
The observations are described in Sect.~\ref{sec-obs}, and analyzed in Sect.~\ref{sec-res}.


\section{Observations}
\label{sec-obs}

The IRAM interferometer is a six-antenna (15 m each) array located on the Plateau de Bure, in the French Alps, and equipped with heterodyne, dual polarization, receivers operating around 0.8, 1.3, 2, and 3 mm (350, 230, 150, and 100 GHz, respectively). 
Garradd was observed  with the PdBI at  wavelengths of 1.1~mm (266 GHz) on 3 March 2012 and 1.9~mm (157~GHz) on 4 March 2012, 70 days after its perihelion passage on 24.6 December 2011. 
Garradd was at  geocentric and heliocentric distances of respectively 1.3~AU and 1.8~AU, and at a phase angle of 31$^{\circ}$.
The comet was tracked using an ephemeris  computed on the basis of the JPL\#75 solution for its orbital elements  (JPL ephemeris are accessible on line at http://ssd.jpl.nasa.gov/horizons.cgi).

The PdBI wide band  correlator (WIDEX) was used to observe the continuum thermal emission of the comet over a total bandwidth of 3.6~GHz in two orthogonal polarizations. 
The array was in a moderately extended configuration  with baseline lengths ranging from 90 to 450~m, providing an elliptical  synthesized beam
with sizes of  0.4$'' \times 1.4''$ at 266 GHz and  0.85$'' \times 0.98''$ at 157~GHz. 
The entire calibration process was performed using the GILDAS software package developed by IRAM \citep{pet05}.

\subsection{Observations at 1.1~mm}
The observations were carried out on 3 March 2012 between 8 and 16~h UT under poor weather conditions  with a phase noise ranging from 40 to 90$^{\circ}$, depending on the baseline length, and a system temperature of the order of 500~K.
These data were unfortunately useless for our purpose.

\subsection{Observations at 1.9~mm}

   \begin{figure}
   \centering
   \includegraphics[width=\columnwidth]{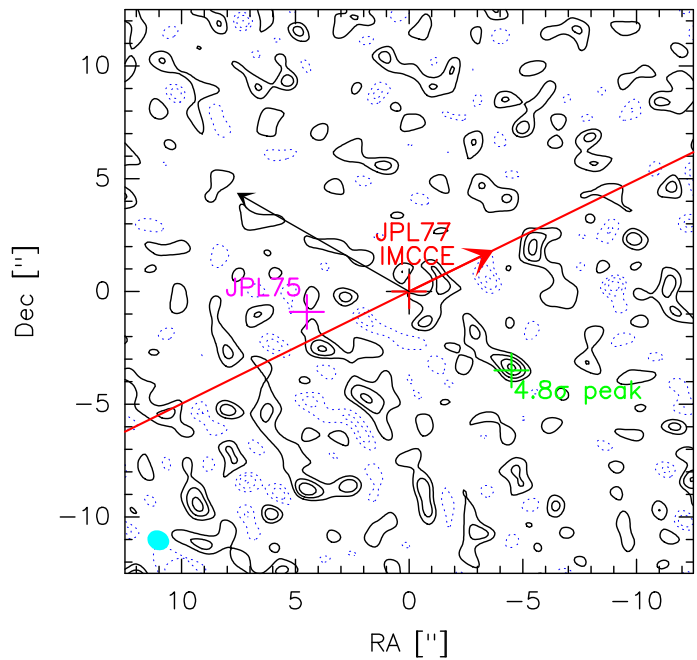}
      \caption{Interferometric map obtained at 1.9~mm on 4 March 2012 with the Plateau de Bure interferometer.
        The map is centred at the expected position of the comet according to the identical ephemeris solutions JPL\#77 and IMCCE (red cross).
         The trajectory of the comet around the observing time is indicated by the red line, the arrow indicates the direction of the motion.
        The tracked position given by JPL\#75 is also indicated (pink cross).
        The emission peak with a SNR of 4.8 (see text) is indicated by the green cross.
       	The arrow points to the direction to the Sun and the ellipse at the bottom left corner represents the synthesized beam.
        The contour spacing corresponds to the  rms value of the signal (0.13~mJy) and the dotted contours represent negative values.
      }
     \label{fig-map}
   \end{figure}

These observations were carried out between 3 March 2012, 22~h UT,  and 4 March 2012, 8~h UT.
The atmospheric conditions were good, with a phase noise in the range 30--65$^{\circ}${} and a system temperature of $\sim$150~K.
The instrumental and atmopheric variations of the phase and amplitude were calibrated using regular observations of the quasars 1435+638 and 1642+690, every $\sim$30~min. 
The absolute flux scale was determined observing  MWC349, with a precision of 10\%.

Given the geocentric distance of the comet (1.3~AU), the synthesized beam of $\sim$0.9$''$ corresponds to a projected distance of $\sim$850~km at the comet. 
The nucleus thermal  emission is thus expected to appear as a point source.
We fitted the Fourier transform of a point source to the observed visibilities and found  an emission peak  9.4$''$ away from the tracked position with a flux of  0.62~$\pm$~0.13~mJy.
This corresponds to a signal-to-noise ratio (SNR) of 4.8, sligthly below the usual limit for a robust detection (SNR = 5). 
This is the only peak with a SNR exceeding 4 in a 20$''$ region around the expected comet position.
The uncertainty on the  absolute flux calibration is  10\%.
We estimated the overall $\pm 1 \sigma$ limits as $F^{+1\sigma} = 1.1 \times (F + 1\sigma) = 0.83$ mJy and $F^{-1\sigma} = 0.9 \times (F - 1\sigma) = 0.44$ mJy.
From this, we deduced new uncertainties on the flux: $F = 0.62^{+0.21}_{-0.18}$ mJy.
Since the emission peak is offset from the pointing centre, the measured flux has to be corrected from the primary beam attenuation (gaussian profile with HPBW of 30.5$''$).
At  a distance of 9.4$''$ from the pointing direction, this attenuation reaches 23\%.
The  corrected flux is then $F = 0.81^{+0.27}_{-0.23}$ mJy.
The interferometric map is presented in Fig.~\ref{fig-map}.
Table~\ref{tab-flux} summarizes the measurements.

\begin{figure}
   \centering
   \includegraphics[width=\columnwidth]{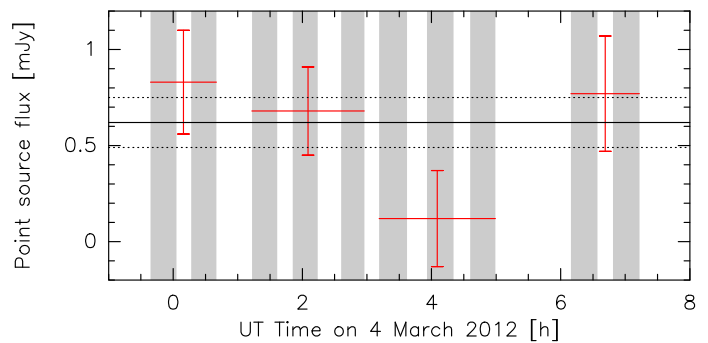}
      \caption{
 Point source fluxes as measured by  fitting  a point source to four visibility subsets. 
The grey areas indicate the periods when the instrument was actually observing the comet. 
The horizontal red bars represent the time coverage of each of the 4  subsets.   
The black solid line represents the flux measured on the entire dataset, the $\pm$1$\sigma$ levels being indicated by the dotted lines.}
     \label{fig-t}
   \end{figure}

This emission peak cannot be attributed to a galactic or extragalactic source since the tracked position moved by more than 15~arcmin in the equatorial frame during the 8-h observing period.
Furthermore, we splitted our visibility set in  four parts and fitted in each of them the Fourier transform of a point source fixed at the position found using the global dataset.
The result is illustrated in Fig.~\ref{fig-t}.
The point source flux found in all the  subsets is consistent  with the overal flux at the 2$\sigma$ level.
 Similar consistancy is obtained when dividing the data in three subsets instead of four.
This confirms that the emission peak we measured is not due to a short-lived interference or noise peak during the observing period.

\begin{table*}
\begin{center}
\caption{Continuum point source detection properties}
\label{tab-flux}
\begin{tabular}{l  c c c c c c c c}
\hline
\noalign{\smallskip}
Day in March & Frequency & RA$^a$ & Dec$^a$      & Obs-JPL\#75$^b$   & Obs-JPL\#77$^c$ & Obs-IMCCE$^d$ & Fitted Flux$^e$    & Corrected Flux$^f$ \\
UT   & GHz       & h:min:s     &  $^{\circ}$:$'$:$''$ &  ($''$,$''$)   &    ($''$,$''$)   & ($''$,$''$)  & mJy        & mJy \\
\hline
\noalign{\smallskip}
\hline
\noalign{\smallskip}
3.917 -- 4.333  & 157.2 &  14:46:22.870  &     $+$68:16:22.38  & (--9,--2.6)   & (--4.4,--3.3) & (--4.4,--3.4)   &  0.62 $\pm$ 0.13 & 0.81$^{+0.27}_{-0.23}$ \\
\noalign{\smallskip}
\hline
\noalign{\smallskip}
\end{tabular}
\end{center}
$^a$ Apparent EQ 2000 positions of the 4.8$\sigma$ brightness peak at the reference time 7h00 UT on 4 March 2012.\\
$^b$ Offset (RA,Dec) in arcsecond between the observed continuum emission peak and the position derived from the JPL\#75 solution for the comet's orbital elements, which is based on astrometric measurements between  August 2009 and January 2012. 
This is the solution which was used to track the comet.\\
$^c$ Offset (RA,Dec) in arcsecond between the observed continuum emission peak and the position derived from the JPL\#77 solution for the comet's orbital elements. 
This solution  takes into account astrometric measurements between  August 2009 and October 2012.\\
$^d$ Offset (RA,Dec) in arcsecond between the observed continuum emission peak and the position derived from the IMCCE solution for the comet's orbital elements. This solution  takes into account astrometric measurements between  August 2009 and January 2013.\\
$^e$ Flux obtained by fitting  the Fourier transform of a point source to the observed visibilities. 
The error bars do not take into account uncertainties in the absolute flux calibration.\\
$^f$ Continuum flux corrected for the primary beam  attenuation (gaussian profile with HPBW of 30.5$''$). 
The error bars take into account  a 10\% uncertainty in the  flux calibration. \\

\end{table*}

\section{Results}
\label{sec-res}
\subsection{Comet position}
\label{sec-pos}
Given the distance between the 4.8$\sigma$ brightness peak and the comet expected position, the question naturally arises as to whether or not this signal can be attributed to the comet.
The JPL\#75 solution, used for comet tracking, was based on astrometric measurements performed between  August 2009 and January 2012, before perihelion.
More recent solutions which include a much larger data set, especially after perihelion,  should be more accurate for March 2012.
We now introduce the solution  JPL\#77 including data up to October 2012 and the solution provided by the Institut de m\'ecanique c\'eleste et de calcul d'\'eph\'em\'erides (IMCCE) with data up to January 2013. 
The two ephemerides predict nearly identical positions (separated by less than $\sim$0.1$''$) slightly closer to the brightness peak detected in the interferometric map (see Fig.~\ref{fig-map} and  Table~\ref{tab-flux}) than that given by the JPL\#75 solution.
 However, the separation remains large, 5.5$''$, corresponding to $\sim$5000 km at the distance of the comet, exceeding by far the  uncertainties related to the ephemeris and the peak position.
Indeed, the (O-C) rms values of the JPL and the IMCCE solutions respectively amount to 0.5$''$ and 0.4$''$, and the precision on the position of a brightness peak measured in interferometric data is of the order of the synthesized beam size divided by the signal-to-noise ratio of the detected flux, yielding $\sim$0.2$''$ in our case.
The distance between our brightness peak and the ephemeris position of the comet (5.5$''$) is thus well above the separation of 2.1$''$ ((0.5$''$+0.2$''$)$\times$3)  which would correspond to an agreement at the 3$\sigma$ level.
Moreover, from our 15--19 Feb. 2012 single dish observations with the IRAM-30m telescope, coarse maps of HCN(3-2) and other species suggested that the peak intensity was within uncertainty ($\sim$1$''$) of the predicted position from the latest orbital solution \citep[JPL\#77,][]{biv+12dps}.

In summary, the 5.5$''$  offset between our brightness peak and the predicted position of the comet seems to be too large to be interpreted in terms of ephemeris and pointing uncertainties or even coma morphology, as it was done in the case of comet C/1995 O1 (Hale-Bopp) for which offsets of the order of 1$''$ \citep{boi+07} were found.

The hypothesis of a technical, systematic problem at the observatory leading to wrong positions is extremely unlikely since previous cometary observations provided excellent agreement between the ephemeris position and the continuum emission detected with the PdBI (e.g., observations of 103P/Hartley 2, Boissier et al., in preparation).

Whereas we do not have an explanation for the brightness peak, we cannot reasonably associate it to comet Garradd.
Considering that the nucleus should have been at the common IMCCE--JPL\#77 position, we can only derive a 3$\sigma$ upper limit on the comet flux of 0.41~mJy at this position.

\subsection{Nucleus size}
\label{sec-size}
We used a standard thermal model described in  previous works \citep{gro+04,lam+10lut} to derive an upper limit for the size of the nucleus assuming  a  spherical shape.
We adopted standard values for the model parameters: null thermal inertia \citep{gro+09,boi+11}, beaming factor $\eta = 1$ \citep{lebspe89}, and  millimetric emissivity $\epsilon = 0.9$.
The  upper limit measured at the expected nucleus position of 0.41 mJy 
corresponds to a nucleus radius smaller than 5.6~km. 
As a side remark, a nucleus radius of 7.8~$\pm$~1.2~km would be required to produce the 0.81~mJy brightness peak found in the data at 5.5$''$ from the comet expected position.
These size estimates do not take into account the dust contribution to the continuum emission as we found it to be negligible (see Sect.~\ref{sec-dust}). 

In order to check whether these sizes are reasonable, we considered the case of a spherical nucleus made of pure water ice and computed its radius required to release the observed amount of water ($\sim$2~$\times$~$10^{29}$~s$^{-1}$) measured  at 1.9~AU with Herschel space telescope \citep{boc+12} and with the Nan\c cay ratiotelescope (\citealt{col+11}, and in preparation).
To do so, we used our standard thermal model to compute the surface temperature and the water sublimation rate integrated over the entire nucleus at this heliocentric distance and found a radius of  4.0~km.
However,  if the active areas are limited to only 30\% of the nucleus surface, the derived nucleus  radius increases to 7.3~km. 
 Given our upper limit on the nucleus radius of 5.6~km, we estimated that the active areas on  Garradd's nucleus cover at least 50\% of the surface, making this comet a very active one.
Admittedly, we do not consider that a significant fraction of the outgassing could be from icy grains, as observed for instance for 103P/Hartley 2 \citep{ahe+11}.

\subsection{Dust contribution}
\label{sec-dust}

The thermal emission of a comet is the sum of the contributions arising from  the nucleus  and the dust grains in the coma. 
Our upper limit on the nucleus size did not consider the contribution of dust thermal emission to the millimetric flux.

In a first approach, the  contribution of the dust coma can be extrapolated from past measurements of the 2~mm continuum emission of comet C/1996 B
(Hyakutake) performed at the James Clerk Maxwell Telescope (JCMT) on 23--24 March 1996 \citep{jewmat97}.
Correcting for the geocentric and heliocentric distances and beam size, the dust flux expected for comet Garradd in the synthesized PdBI beam
would have amounted to 0.22~mJy. 
One must further consider differences in the activity of the two comets. 
At the time of the JCMT observations, comet Hyakutake displayed a water production rate of $Q$(H$_2$O) $\sim$ 2 $\times$ 10$^{29}$ s$^{-1}$ and $Af\rho$ $\sim$ 7000 cm \citep{schosi02}. 
The values for comet Garradd on 14 March 2012 are $Q$(H$_2$O)~=~4.4~$\times$~10$^{28}$~s$^{-1}$ and $Af\rho$~=~3400~cm (Schleicher 2012, personal communication). 
\cite{far+12} reported  very similar values with  $Q$(OH)~=~1.9~$\times$~10$^{28}$~s$^{-1}$ and $Af\rho$~=~3400~cm on 6 March 2012 measured with the Medium Resolution Instrument (MRI) onboard the Deep Impact spacecraft.
We note that these $Af\rho$ values were obtained in the same spectral domain (visible) and that the two comets were at comparable phase angles so that the comparison is not biased.
The water production rate and $Af\rho$ are thus respectively $\sim$5 and $\sim$2 times lower than for comet Hyakutake.
Hence, the dust contribution in the PdBI beam is not expected to exceed 0.1~mJy, which is lower than the uncertainty of the PdBI measurements,
unless comet Garradd's nucleus was releasing comparatively bigger particles.

In a second approach, it is possible to assess how the dust emission may affect the signal by determining the dust production rate required to explain the observed flux. 
We used the model of dust thermal emission presented by \cite{boi+12}.
Absorption cross-sections were calculated with the Mie theory, taking a refractive index (($n$,~$k$)~=~(2.05, 0.007) at 1.91~mm) corresponding to porous grains ($P$~=~0.5) composed of a 50:50 mixture of crystalline  \citep{fab+01} and amorphous silicates \citep{dra85}.
We also considered pure organic grains with refractive index ($n$,~$k$)~=~(2.28,~0.0028) of \cite{pol+94}.
As introduced by \cite{newspi85} and \cite{han+85IRAS,han+85CG}, and extensively used thereafter (e.g., \cite{tot+05}), we considered a differential dust production $Q_{\rm dust}$($a$) as a function of grain radius $a$, described by the size index $\alpha_d$, between --4 and --2.5. 
We assumed a minimum grain size of $a_{min}$ = 0.1 $\mu$m.
The maximum grain size $a_{max}$, as well as size--dependent velocities $v_{\rm d}$($a$) were computed following \cite{crirod97}. 
We assumed the same bulk density of 500 kg~m$^{-3}$ for the nucleus and the dust grains  and  considered two cases for the size of the nucleus: 
a relatively small one with  $r_N$ = 2~km (model 1) and a larger one with $r_N$ = 6~km, similar to the upper limit on the nucleus size estimated in  Sect.~\ref{sec-size} (model 2).
Table~\ref{tab-mod} presents the grain maximum sizes and  velocity ranges, as well as the dust production rates required to produce a flux of 0.41~mJy, corresponding to our upper limit on the comet emission.

The derived dust production rates (ranging from 0.5 to 4~$\times$~10$^{4}$~kg~s$^{-1}$) widely exceed (by factors of 4 to 30) the gas production rates of 1300~kg~s$^{-1}$, considering only water production (Schleicher 2012, personal communication). 
Since comets display dust-to-gas mass ratios lower than 5  \citep[see e.g.,][for measurements in the dusty C/1995 O1 Hale-Bopp]{wei+03},
the actual dust production rate is likely to be much less than our estimates.
As a consequence, we consider that the dust contribution to our signal is negligible, and that the upper limit of 5.6~km derived previously for the nucleus is robust.

\begin{table}
\caption{Dust coma model parameters and production rate upper limits.}
\label{tab-mod}
\begin{tabular}{c l l c c} 
\hline
\noalign{\smallskip}
& Model number& & 1 & 2 \\ 
\hline
\noalign{\smallskip}
\hline
\noalign{\smallskip}
$r_N$ & Nucleus radius & km & 2 & 6 \\
$a_{min}$ &  Smallest grain size & $\mu$m & 0.1 &   0.1        \\
$a_{max}$ &  Largest grain size & cm & 13 &   0.48        \\
$v_d$    & Velocity range & m~s$^{-1}$ & 1.6--490& 4.8--406  \\
$Q_d$$^{\ast}$    & Silicatic grains & kg s$^{-1}$ &1.6--2.4~$\times$~10$^{4}$  & 0.5--2.1~$\times$~10$^{4}$ \\
$Q_d$$^{\ast}$   & Organic grains   &  kg s$^{-1}$ & 1.9--4.0~$\times$~10$^{4}$& 1.1--4.3~$\times$~10$^{4}$\\
\hline
\noalign{\smallskip}
\end{tabular}

$^{\ast}$ Dust production rate for a flux of 0.41 mJy at 1.9 mm. The range  corresponds to size indexes $\alpha_d$ between --4 and --2.5.
\end{table}

 This calculation further allows us to rule out the 4.8$\sigma$ brightness peak as resulting from a hypothetical dust outburst at an earlier date. This peak of 0.8 mJy translates in a dust production rate of 1--9~$\times$~10$^{4}$  kg s$^{-1}$  well above what is reasonably conceivable unless a major event took place such as the fragmentation of the nucleus. Indeed fragmentation events in comets C/1996 B2 (Hyakutake) and 73P/Schwassmann-Wachmann 3 produced dust stuctures in the antisolar direction observed in optical images \citep[e.g.,][]{ish+09,vin+10}.
However the magnitude of the hypothetical dust outburst is such that it should have had an optical counterpart, for instance a strong increase of the comet visual brightness resulting from the associated production of small grains reflecting sunlight. 
No such outburst was reported \citep[e.g.,][]{nicim12a,nicim12b} casting very serious doubt on this scenario. 
Moreover, large grains prominently contributing to millimetre emission are weakly affected by solar pressure and would not spread in the antisolar direction -- contrary to the  4.8$\sigma$ peak -- but rather be distributed along the projected  comet trajectory (very much like a cometary dust trail) which happened to be almost orthogonal to the antisolar direction  (see Fig.~\ref{fig-map}).

\section{Summary}

We observed the thermal emission of  the comet C/2009 P1 (Garradd) at 157~GHz (1.9~mm) using the IRAM Plateau de Bure interferometer in March 2012.
A brightness peak of 0.81$^{+0.27}_{-0.23}$~mJy was detected at a distance of 5.5$''$ from the position provided by the solutions IMCCE--JPL\#77 for the comet's orbital elements.
Such an offset ($\sim$5000~km at the comet) is too large to be interpreted in terms of coma morphology and it is thus unlikely that this emission is coming from the comet.
We then derived an upper limit on the comet thermal emission of 0.41 mJy at the position of the comet provided by the ephemeris.
Two different approaches were implemented to ascertain that the dust coma contribution to the millimetric emission in the PdBI synthesized beam is negligible.
We therefore established a robust 3$\sigma$  upper limit on the nucleus radius of 5.6~km.
According to the compilation of \cite{lam+04}, 8 out of 13 NICs for which we have reliable estimates have nucleus radius less than 5.6~km.
Although the statistics is limited, it appears that the nucleus C/2009 P1 (Garradd) has a size typical of NICs.
 Provided  the production rates of this comet measured independently from our study, we showed using a simple model of comet activity that Garradd is a very active comet, with an active fraction of the nucleus of at least 50\%.

\begin{acknowledgements}
The research leading to these results  received funding from the European Community's Seventh Framework Programme (FP7/2007--2013) under grant agreement No. 229517.
We thank P. Rocher from IMCCE for providing us with the ephemeris of the comet.
\end{acknowledgements}

\bibliography{}

\end{document}